\documentstyle[aps,epsf]{revtex}
\tighten
\begin{document}
\draft

\title{Random Resonators and Prelocalized Modes in Disordered 
Dielectric Films.}

\author{V. M. Apalkov and M. E. Raikh}
\address{Department of Physics, University of Utah, Salt Lake City,
UT 84112, USA}
\author{B. Shapiro}
\address{Department of Physics, Technion-Israel Institute of
Technology, Haifa 32000, Israel}
\maketitle

\begin{abstract}
Areal density of disorder-induced resonators with a high quality factor,
$Q\gg 1$, in a film with fluctuating refraction index is calculated
theoretically. We demonstrate that for a given $kl>1$, where $k$ is
the light wave vector, and $l$ is the transport mean free path, when
 {\em on average} the light propagation is diffusive, the likelihood 
for finding a random resonator increases dramatically with 
increasing the
correlation radius of the disorder. Parameters of {\em most probable}
resonators as functions of $Q$ and $kl$ are found. 
\end{abstract}

{\em Introduction}. Recent discovery of the {\em coherent} lasing from
various disordered materials adds a new dimension to the conventional
physics of light propagation in multiply scattering media.  In the
experimental works\cite{cao99,cao00,Fro99,Fro99a} it was demonstrated
that, above a certain excitation level, the emission spectra of {\em
ZnO} powders\cite{cao99,cao00}, conjugated polymer films\cite{Fro99}
and dye-infiltrated opals\cite{Fro99a} exhibit a sequence of extremely
narrow peaks, their widths being limited by the spectrometer
resolution.  Until recently, this finite resolution left some room for
doubt as to whether the observed peaks indicated true
lasing\cite{Wiersma2000}. However, the latest experiments on photon
statistics\cite{cao01,Polson01} have unambiguously established the
coherence of the emitted light, thus proving conclusively that the
underlying mechanism of random lasing in
Refs.~\onlinecite{cao99,cao00,Fro99,Fro99a}
involves the {\em amplitude} (coherent) 
rather than {\em power} (incoherent) feedback. (The latter is known
to occur via the diffusion process of the light intensity,
as proposed long ago by
Letokhov\cite{letokhov} and observed first by Lawandy\cite{Law94} and
later in other numerous experiments.)
The origin of coherent feedback, responsible for the lasing observed
in Refs. \onlinecite{cao99,cao00,Fro99,Fro99a}, is a subject of 
controversy and debate.
Such a feedback would naturally emerge if the light
were localized. However, the coherent backscattering measurements
\cite{cao99,cao99'',cao99',soest,Polson'01,thesis} 
carried out in parallel with the analysis of the emission spectra  
rule out this possibility. Firstly, the values of $kl$ extracted
from these measurements turn out rather large. Secondly, the onset of
the Anderson localization manifests itself in the rounding of the top
of backscattering cone  
\cite{Schmuurmans99,Berkovits87,edrei90}. No such rounding was observed
in the experiments\cite{cao99,cao99'',cao99',soest,Polson'01,thesis}.

In the early work of Cao {\em et. al.}\cite{cao99,cao99'',cao99'}, 
it was argued that, even in diffusive regime of light propagation,
a photon, scattered off a certain sequence of impurities
(grains of, roughly, $100nm$ size) can move on a closed loop, which
thereby serves as a laser resonator. The picture of such ring cavities
put forward in Refs.\onlinecite{cao99,cao99'',cao99'}
was based on  the experimental evidence\cite{wiersma1}
that the {\em recurrent} scattering processes contribute to
backscattering albedo.
However, since in each scattering act most of
the energy gets scattered out of the loop, an unrealistically
high gain would be required to achieve the lasing threshold condition
for such a loop. This point was particularly emphasized 
by van Soest\cite{thesis}, who also argued that 
"impurity loops" are likely to generate
a broad frequency spectrum rather than isolated resonances.

Certainly the picture of random cavities representing a certain
spatial arrangement of {\em isolated} scatterers is too naive.  This,
however, does not rule out the entire concept of disorder-induced
resonators. Although  sparse, the disorder configurations 
that trap the light
for long enough time can occur in a sample of a large enough size, and
a {\em single} such configuration is already sufficient for lasing to
occur.  Therefore, under the condition $kl \gg 1$, which implies that
{\em overall} scattering is weak, the conclusion about the relevance of
random cavities can be drawn only upon the {\em quantitative}
calculation of their likelihood.
This is the subject of the present paper.

We start with a remark  that the issue of random
resonators for light waves has its counterpart in the electronic 
transport. In particular, the modes with anomalously low losses 
are analogous to the so called {\em prelocalized} electronic states
in diffusive conductors that are responsible for the long-time
asymptotics of the current relaxation.
 Theoretical study of these states  was  launched more 
than a decade ago  (see Ref. \onlinecite{Kravtsov}) and 
later renewed in Ref. \onlinecite{Muzykantskii95}. 
The results obtained to date are summarized in the 
review\cite{mirlin00}. Following this analogy we dub the modes
of  
random resonators with anomalously high quality factor, $Q$, as 
{\em prelocalized} modes.

The principal outcome of our study is that for a given $kl \gg 1$, the
probability of formation of a high-$Q$ random resonator depends
crucially on the {\em size} of the scatterers, or, more precisely, on
the correlation radius of the disorder, $R_c $.  
Similarly to the treatment in  
Refs.~\onlinecite{cao00,cao99'} we restrict our consideration to
the two-dimensional case (a disordered film).
Regarding the geometry of a random resonator, we adopt 
the idea  proposed by Karpov in
Ref.~\onlinecite{karpov93} for trapping the  acoustic waves
in three dimensions. According to Ref.~\onlinecite{karpov93}, the 
fluctuations responsible for trapping are the toroidal inclusions
with reduced  sound velocity. 
Correspondingly, in two dimensions, a  random resonator represents 
a ring-shape area  (see Fig.~1a) 
within which the  dielectric constant 
is  enhanced  by some small value $\epsilon _1$ 
(compared to the background value $\epsilon _0$). Then
such a ring can be viewed as a waveguide that supports the modes of
a whispering-gallery type. Due to the azimuthal symmetry, these modes
are characterized by the radial index and the angular momentum,
$m$. Denote by ${\cal N}_m(kl,Q)$ the areal density of resonators
with quality factor $Q$ in the film with a transport mean free path
$l$.  Here $k= \epsilon_0^{1/2} k_0$, where $k_0 = 2\pi /\lambda $,
$\lambda $ stands for the wavelength in vacuum. 
Obviously, in the diffusive
regime, $kl > 1$, the density ${\cal N}_m(kl,Q)$ is exponentially 
small for $Q \gg 1$. In this domain ${\cal N}_m(kl,Q)$ can be
 presented as
\begin{equation}
{\cal N}_m(kl,Q) = {\cal N} _0 e^{-S_m(kl, Q)} ~,
\end{equation} 
 where ${\cal N} _0$ is the prefactor. Now we can ``quantify''
the above statement about the dependence of the resonator
likelihood on $R_c$  
\begin{equation}
\frac{ S_m(k_0R_c > 1)}{ S_m(k_0R_c \ll  1)} =
\frac{ \Phi(\epsilon _1^{1/2} k_0 R_c)}
{\pi ^{1/2} (\epsilon _0 ^{1/2} k_0 R_c)^3}~.   \label{eq1}
\end{equation}
The dimensionless single-parameter function $\Phi $ is 
shown in Fig.~2. It is seen from Eq.~(\ref{eq1}) 
that $S_m$ falls off rapidly with increasing $R_c$.
In the domain $k_0 R_c > 1 $ but $\epsilon _1 ^{1/2} k_0 R_c 
\lesssim 1$ we can set $\Phi \approx 1$, so that we have  
 $S_m \propto (k_0 R_c)^{-3}$. For larger $R_c$, when  
$\epsilon _1 ^{1/2} k_0 R_c \gg 1 $, the function $\Phi $ 
behaves as $\Phi (u) \propto u $. In this domain 
$S_m$ decreases slower with $R_c $: $S_m \propto (k_0 R_c)^{-2}$.
We emphasize again that Eq.~(\ref{eq1}) applies {\em for a given}
$kl$ value, so that the decrease of $S_m$ with $R_c$ {\em leaves 
the backscattering cone unchanged}. Below we will demonstrate 
with rigorous calculations  that the value of 
$S_m(k_0R_c \ll 1)$ is really large
\begin{equation}
S_m (k_0 R_c \ll 1)  
   =  2\left( \frac{\pi ^3 }{3 } \right)^{1/2} kl  |\ln Q |
              ~, \label{eq2}
\end{equation}
which makes 
formation of random resonators with 
$Q\gg 1$ practically impossible for point-like scatterers.   
On the contrary, for $k_0 R_c \gtrsim 2 $ ring-like fluctuations 
of the dielectric constant (see Fig.~1b) make  realistic 
candidates for high-$Q$ random resonators. 

{\em Derivation of Eq.~(\ref{eq1})}. Denote by $P$ the probability
  of fluctuation of dielectric constant, $\delta \epsilon (\bbox{r})$.
Assuming the fluctuations to be Gaussian with r.m.s. $\Delta $, we have 
\begin{equation}
\ln P =
    - \frac{1}{2 \Delta ^2} \int \! \! \! \int d \bbox{r} 
 d\bbox{r}^{\prime }
    \delta \epsilon (\bbox{r}) \delta \epsilon (\bbox{r}^{\prime }) 
 \kappa (\bbox{r}- \bbox{r}^{\prime })  \label{eq3},
\end{equation}
where the kernel $\kappa(\bbox{r})$ is related in the usual way, 
$\int d\bbox{r}^{\prime } \kappa (\bbox{r}- \bbox{r}^{\prime })
 K (\bbox{r} ^{\prime } - \bbox{r}_1) = \delta (\bbox{r}- \bbox{r}_1)$,
to the correlation function, $K(\bbox{r})$, defined as    
\begin{equation}
\langle \delta \epsilon (\bbox{r}) \delta \epsilon 
(\bbox{r} ^{\prime }) \rangle = \Delta ^2
              K (\bbox{r}- \bbox{r}^{\prime }) ~. \label{eq4}
\end{equation}
We search for fluctuations of a type shown in Fig.~1, {\em i.e.},
$\delta \epsilon $ is  azimuthally symmetric (depends only on 
the radius, $\rho $) and is non-zero 
within the relatively narrow ring of width $w \ll \rho _0$, where
$\rho _0$ is the radius corresponding to the middle of the ring
(see Fig.~1a).  
For such a fluctuation the wave  equation for field distribution 
corresponding to the  angular momentum, $m$, reduces to 
\begin{equation}
\hat{L}\chi _m=
\frac{d^2 \chi_m}{d x ^2} + \delta \epsilon ~ k_0^2 \chi_m =
         \epsilon _1 ~k_0^2 \chi _m ~,
\label{eq5}
\end{equation}
where $x=\rho -\rho _0$. The ``eigenvalue'' in the
r.h.s. of Eq.~(\ref{eq5}) is defined as 
$\epsilon _1 = \left(\frac{m}{k_0 \rho_0 } \right)^2- \epsilon _0 $.
It is independent of $x$ by virtue of the condition $\rho _0 \gg w$.
The next step is to find the ``most probable'' distribution $\delta
\epsilon (x)$ for a given $\epsilon _1$. This is done within the 
standard optimal fluctuation approach \cite{Halperin66,Zittartz66},
which prescribes minimization of auxiliary functional $\Psi \{\delta
\epsilon \} = \ln P - \lambda (\chi _m \hat{L} \chi _m)$, where
$\lambda $ is the Lagrange multiplier. This yields
\begin{equation}
\delta \epsilon (x ) = \int dx_1  K_0 (x-x_1) \chi _m^2(x_1 )~,
  \label{eq6}
\end{equation}
where we have set $\lambda = 2\pi \rho _0 /\Delta ^2 k_0^2 $. 
The above expression reminds the standard result of optimal 
fluctuation approach for a correlated random potential 
\cite{Halperin66}. The only difference is that,  
due to the angular integration in Eq.~(\ref{eq3}), 
the kernel in Eq.~({\ref{eq6}) is given by a 
function $K_0 $, which is related to the correlator, 
$K(\bbox{r})$, as  $K_0 (x_1-x_2) = \int_{-\infty }^{\infty } dy 
K (\sqrt{ (x_1-x_2)^2 + y^2})$.
A natural $x$-scale for  the eigenmode $\chi _m$ is $(\epsilon_1 ^{1/2}
k_0)^{-1}$. Thus, we present the outcome of the optimal fluctuation 
approach in terms of a dimensionless variable 
$z=\epsilon _1^{1/2} k_0 x$. 
\begin{equation}
S_m = \frac{\pi \rho _0}{\epsilon _1 k_0^2 \Delta ^2} 
\int \! \! \! \int dz_1 dz_2
\chi_m ^2 (z_1) \chi _m^2 (z_2) K_0 (z_1-z_2)~.
   \label{eqP}
\end{equation}
The dimensionless equation for the function $\chi _m$ 
reads
\begin{equation}
\frac{d^2 \chi _m(z)}{dz^2} +\frac{ \chi _m(z)}{\epsilon _1^{3/2} k_0 }
 \int dz _1 K_0(z-z_1) \chi _m^2(z_1)  =
     \chi _m(z) ~.
  \label{eqChi}
\end{equation}
To proceed further we have to specify the correlator, $K$.  
We have chosen the Gaussian form  
$K(\rho ) = \exp \left( - \rho^2 / R_c^2 \right)$. 
The form of the function $\chi _m = A(R_c) \exp[-\gamma (R_c) z^2]$
allows to cover the entire range of correlation radii, from 
``white noise'' ($R_c \rightarrow 0$) 
to the limit of a smooth disorder ($k_0 R_c \gg 1$). Indeed, for 
large $R_c$ this form becomes exact. In the opposite limit,
$R_c \rightarrow 0$, using the above trial function instead of
exact solution $\chi _m \propto 1/\cosh (z)$ leads to 
the overestimate of $S_m$ by a factor $(\pi /3)^{1/2}\approx 1.023$.
The parameter $A(R_c)$ and $\gamma (R_c)$ of the trial
function can be found analytically. Evaluation of  Eq.~(\ref{eqP}) 
reduces to the Gaussian integration which for a given $m \approx 
\epsilon ^{1/2} k_0 \rho _0$ yields 
\begin{equation}
S_m =2^4 3^{-3/2} \pi ^{1/2} m 
\left(\frac{\epsilon _1^3}{\epsilon _0 }\right)^{1/2}
\frac{\Phi (\epsilon _1^{1/2}k_0R_c )}
{(\Delta  k_0 R_c)^2} ~.
   \label{eq7}
\end{equation}
The analytical expression for  the function $\Phi (u)$ introduced 
above in Eq.~(\ref{eq1}), and shown in Fig.~2, is the following 
\begin{equation}
\Phi(u)  =
\frac{3^{3/2}}{16} 
\left( \frac{5+ \sqrt{9+16 u^2}}{3+ \sqrt{9+16 u^2}}\right)^2
   \left[ u^2 + \frac{1}{2} ( 3+ \sqrt{9+16 u^2}) \right]^{1/2} ~.
  \label{eq8}
\end{equation}
Recall that we are interested in the density of random resonators at a
{\em given } value of $kl$. The remaining task is to express the
transport mean free path in terms of $\Delta $ and $R_c $. With
$K(\rho ) = \exp \left( - \rho^2 / R_c^2 \right)$ the expression
simplifies in two limits
\begin{mathletters}
\begin{eqnarray}
& & kl =  \frac{4 \epsilon _0 }{\pi (k_0 R_c\Delta )^2}  ~~\mbox{for}
~~k_0R_c \ll 1~;   \label{eq9a} \\
& & kl =  \frac{4\epsilon _0^{5/2} k_0R_c}{\pi ^{1/2} \Delta ^2 }
~~\mbox{for}~~~k_0R_c \gg 1 ~\label{eq9b} ~.
\end{eqnarray}
\end{mathletters}
Note, that Eqs.~(\ref{eq9a}), (\ref{eq9b}) can be cast into the 
conventional form $l\propto 1/n\sigma $, where $n$ is the concentration 
of scatterers and $\sigma $ is the transport cross section. In 
particular, Eq.~(\ref{eq9a}) follows from the two-dimensional 
version ($\sigma \propto R_c^4 k_0^3$) of the Rayleigh scattering 
formula for $\sigma $. 
Combining Eq.~(\ref{eq7}) and Eqs.~(\ref{eq9a}), (\ref{eq9b}), 
we arrive at 
 Eq.~(\ref{eq1}). In the limit $\epsilon _1^{1/2} k_0 R_c \gg 1$ 
Eq.~(\ref{eq7}) has a simple interpretation. Namely, in this
limit $S_m $ can be rewritten as 
$S_m \sim (\epsilon _1 /\Delta )^2 (\rho _0 /R_c)$.  
The first factor comes from 
the Gaussian probability to have a fluctuation $\epsilon _1$ within 
a square of an area $\sim R_c ^2$. The second factor accounts 
for the number of  squares needed to cover the ring area 
$A_m \sim \rho _0 R_c$.

{\em $Q$-factor}. It is obvious that at large distances, 
$\rho \gg \rho _0$, the behavior of $\chi _m$ is oscillatory,
$\chi_m \propto \exp (i \epsilon _0^{1/2} k_0 \rho )$, manifesting
that the waveguided mode of a ring, being prelocalized, 
has a finite lifetime.
In other words, the eigenvalue, $\epsilon _1$, in Eq.~(\ref{eq5})
has an imaginary part, $\tilde{\epsilon } _1$, due to 
evanescent leakage. The quality factor is inversely proportional 
to $\tilde{\epsilon } _1$, namely, $Q= \epsilon_0 / 
\tilde{\epsilon } _1$. The leading contribution to
$\tilde{\epsilon } _1$ comes from the region of a 
width, $d$, adjacent to the waveguide (see Fig.~1a). To 
find $d$ we have to take into account that the r.h.s in 
Eq.~(\ref{eq5}) does in fact depend on $x$. This is because the 
precise form of the r.h.s is not $\left( \frac{m}{\rho _0}\right)^2 -
\epsilon _0 k_0 ^2$, but rather $\left( \frac{m}{\rho _0+x}\right)^2 -
\epsilon _0 k_0 ^2$. In the region outside the  waveguide, when
$x \gg 1/\epsilon _1^{1/2}k_0$ (but still $x \ll \rho _0$)
Eq.~(\ref{eq5}) takes the form
\begin{equation}
\frac{d^2\chi_m}{dx^2} = 
\epsilon _1 k_0^2 \left( 1- \frac{x}{d} \right) \chi_m
  \label{eq10}~,
\end{equation}
where the width of the decay region is given by
$d = \epsilon _1 k_0^2 \rho_0^3 /2m^2$.
Equation (\ref{eq10}) is of Airy-type. Semiclassical calculations 
with exponential accuracy yields for the rate of evanescent leakage 
$\tilde{\epsilon}_1 \propto \exp [- (2m/3) 
( \epsilon _1 /\epsilon _0 )^{3/2} ]$, and hence 
\begin{equation}
\ln Q     =
   \frac{2m}{3} \left( \frac{\epsilon _1}{\epsilon _0}
          \right)^{3/2} \label{tt} ~~.
\end{equation}
Substituting $\epsilon _1$ from Eq.~(\ref{tt}) and 
$\Delta $ from Eq.~(\ref{eq9a}) into Eq.~(\ref{eq7}) we arrive at 
Eq.~(\ref{eq2}). The above derivation relied on the assumptions 
$\rho _0 \gg d $ and $d\gg w$. These 
assumptions are justified within the following domain of  the 
quality factors $m \gg \ln Q \gg \max \left\{ 1,~ 
\epsilon _0^{3/4} (k_0 R_c )^{3/2} m^{-1/2} \right\}$. 

{\em Discussion.} Equation (\ref{eq2}) quantifies the effectiveness of 
trapping of light in a random medium with point-like scatterers.
It follows from Eq.~(\ref{eq2}) that the likelihood of high-$Q$ 
cavity  is really small. Indeed, even for rather strong 
disorder,  
$kl = 5$,  the exponent, $S_m$,  in the probability of
having a cavity with a quality factor $Q=50$ is close to $S_m = 120$.
We emphasize that in two dimensional case under consideration, 
this exponent does not depend on $m$ and, thus, on the cavity
radius $\rho _0 = m/\epsilon _0^{1/2} k_0$. To estimate 
the degree to which finite size of scatterers ($\sim R_c$) 
improves the situation we choose $k_0 R_c \approx 2$, which 
already corresponds to the limit 
$k_0 R_c \gg 1$ in Eq.~(\ref{eq9b}), 
but still allows to set $\Phi = 1$. Then for $Q=50$, 
$kl = 5$ we obtain 
$S_m \approx  1.1$, suggesting that the resonators 
with this $Q$ are quite frequent. In the latter estimate 
we have set $\epsilon _0 =4$.

A natural question to address is how large a value of 
 $Q$ can be achieved for a given $kl$ and 
$k_0 R_c$. To address this question we inspect the  argument 
$u = \epsilon _1^{1/2} k_0 R_c $ of the  function $\Phi $. 
With the use of Eq.~(\ref{tt}) it can be presented in the form  
$ u = \epsilon _0^{1/2} k_0 R_c 
\left[ \frac{3\ln Q }{2m} \right]^{1/3}$. Since $\Phi (u) $ 
increases monotonically (see Fig.~2), the latter expression 
suggests that $Q$ can be increased at the expense of larger 
$m$-values. In the example considered above, in order to 
keep $\Phi $ smaller than, say $1.5$, $m$ should be bigger than 
$50$. However, due to slow dependence $u\propto m^{-1/3}$, 
we get rather small value $S_m\approx 3$ for $m$ as small as $m=15$. 
Certainly, the allowed values of $m$ are limited from 
above. This limitation originates from ``vulnerability'' of
waveguiding to the fluctuations of the dielectric constant 
around optimal ring-like distribution. The dangerous fluctuations are
those that enhance the evanescent leakage. Note, that these 
fluctuations do not affect the main exponent $S_m$ in the 
density of resonators. It is obvious that the bigger is the 
area $A_m = 2 \pi \rho _0 d\propto m^{4/3}$, responsible for  
evanescent leakage (see Fig.~1a), the harder it is to ``protect'' 
the waveguiding. Since the fluctuations $\delta \epsilon (\bbox{r})$ 
have the spatial scale $R_c$, the probability that the waveguiding 
``survives'' can be roughly estimated as $\exp (-A_m / R_c^2)$. 
The condition that the exponent $ A_m / R_c^2$ does not 
dominate over the principal exponent $S_m$ can be rewritten 
as $m\lesssim  (\epsilon _0 kl )^{3/5} (\ln Q)^{2/5}$. For the example 
$kl = 5$ and $Q = 50$, addressed above, we get $m\lesssim 10 $. 
Rigorous calculation of the ``survival probability'' is the problem 
of the same complexity as calculation of the prefactor in 
the functional integral \cite{Brezin80}.

{\em Conclusion}. 
In the present paper we  provided a quantitative
theory of random resonators that 
 substantiates the intuitive image 
\cite{cao99,cao99''}
of a resonant  cavity as a closed-loop
trajectory of a light wave bouncing between the point-like
scatterers. The  intuitive picture in 
\cite{cao99,cao99''} assumed that light can
propagate along a  loop of scatterers by simply being
scattered from one scatterer to another. Such a picture, however, is
unrealistic due to the scattering out of the loop \cite{thesis}. 
 We have 
demonstrated that the scenario of light traveling along 
closed loops can be remedied. In our picture the "loops", 
i.e. the random resonators, can be envisaged as rings with 
dielectric constant larger than the average value.
On a microscopic level these  resonators
correspond to certain arrangements of scatterers (grains). The main
point, however, is that a resonator acts as a {\em single entity}: 
only
 the coherent multiple scattering of light by {\em all} the scatterers
in the resonator can provide trapping. We also point out that 
correlations in the fluctuating part of the dielectric constant
(due to finite grain size) highly facilitate trapping.


The effectiveness of light trapping is expressed by
Eq.~(\ref{eq1}). This expression describes the statistics of 
the quality factors which determines the distribution 
 of the threshold gain for random
lasing. Our consideration pertains to the {\em passive} disordered
films, in the sense, that we neglect the effect of gain on the spatial
distribution of the light intensity
\cite{cao00,cao99',Jiang00,Jiang01}.  Random resonators considered in
the present paper are sparse, so that there is no spatial overlap
between the modes of different resonators, the situation opposite to
that considered in Refs.  \cite{beenakker96,beenakker98}.  We have
also treated scatterers as frequency-independent fluctuations, $\delta
\epsilon (\bbox{r})$, of the dielectric constant. The entirely
different scenario of the collective mode  
formation emerges for {\em resonant}
scatterers \cite{cooperative}.

{\em Acknowledgments}. Two of the authors MER and BS
acknowledge the hospitality of the University of K\"{o}ln and of
the Institute for Theoretical Physics at UCSB where the parts of
this work were completed.
This research was supported in part by the National Science
Foundation under Grant No. PHY99-07949.

\begin{figure}
\centerline{
\epsfxsize=4.8in
\epsfbox{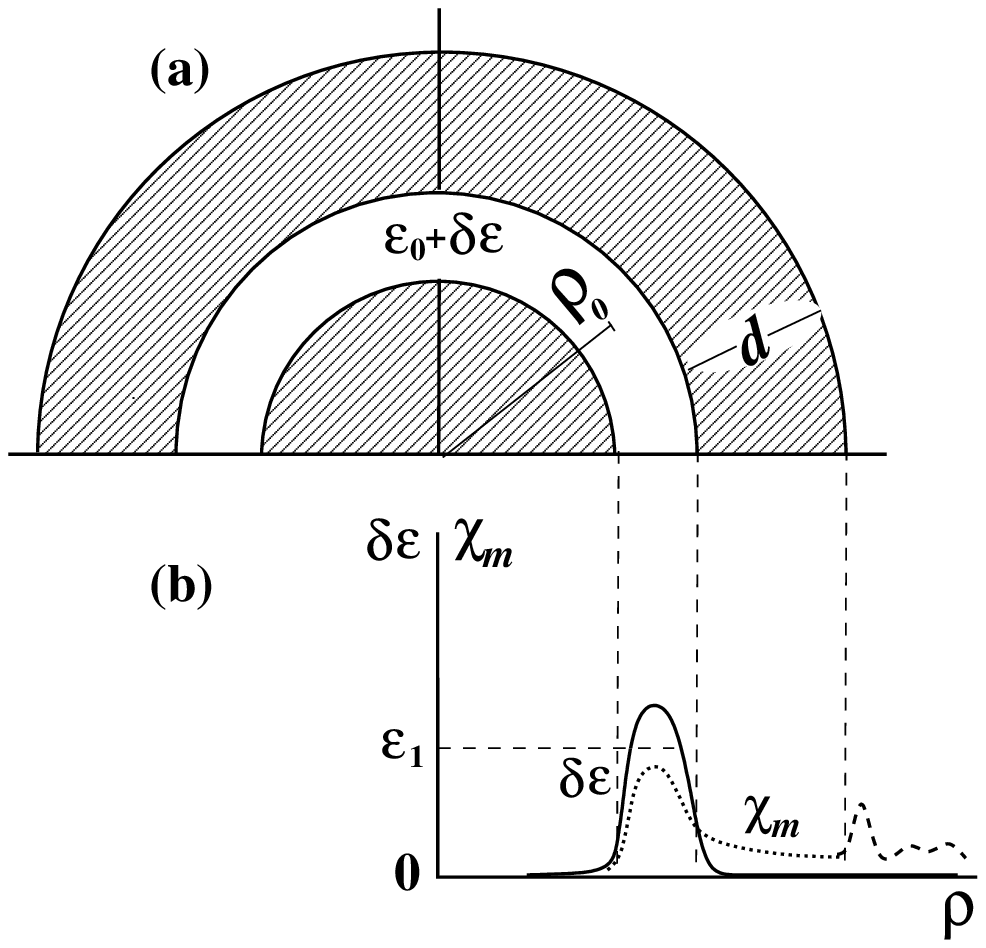}}
\vspace*{0.1in}
\protect\caption[sample]
{\sloppy{(a) The structure of a two-dimensional
resonator is illustrated schematically; only a half of the
ring-shaped waveguide (blank region) is shown. 
(b) Optimal fluctuation of the dielectric constant, $\delta \epsilon 
(\rho )$ (solid line), and the corresponding field distribution 
(dotted line) are shown. Dashed line outside the shaded region 
of a width, $d$, illustrates the evanescent leakage.  
}}
\label{figone}
\end{figure}

\begin{figure}
\centerline{
\epsfxsize=4.8in
\epsfbox{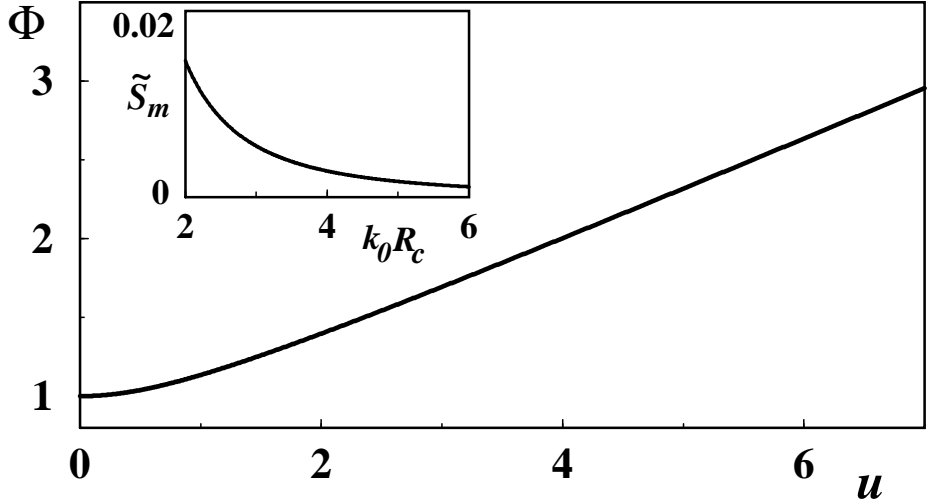}}
\vspace*{0.1in}
\protect\caption[sample1]
{\sloppy{Dimensionless function $\Phi(u)$ defined in Eq.~(\ref{eq8})
is plotted. Inset: Normalized modulus of the log-density of random
resonators, $\tilde {S}_m = S_m(k_0 R_c)/S_m(0)$,
calculated from Eq.~(\ref{eq1}) for  $\epsilon_0 =4$, $Q=50$, 
and $m=15$ is plotted versus the dimensionless 
correlation radius, $k_0 R_c$. 
}}
\label{figtwo}
\end{figure}

\end{document}